\documentclass{aa}
\newcommand{\oma}{\omega_{\rm A}}
\newcommand{\Om}{\Omega}
\newcommand{\be}{\begin{equation}}
\newcommand{\ee}{\end{equation}}
\newcommand{\bea}{\begin{eqnarray}}
\newcommand{\eea}{\end{eqnarray}}
\newcommand{\mk}{}

\begin{document}
\title{Dynamo action by differential rotation in a stably stratified stellar interior}
\titlerunning{Dynamo action in a stellar interior}
\authorrunning{H.C.\ Spruit}
\author{H.C.\ Spruit\inst{1,2}}

\institute{
Institute for Theoretical Physics, University of California, Santa Barbara, CA 93106
\and Max-Planck-Institut f\"ur Astrophysik, Postfach 1317, 85741 Garching, Germany
}

\offprints{henk@mpa-garching.mpg.de}
\date{Received / Accepted}

\abstract{Magnetic fields can be created in stably stratified (non-convective) layers in a
differentially rotating star. A magnetic instability in the toroidal field (wound up by
differential rotation) replaces the role of convection in closing the field amplification loop.
Tayler instability is likely to be the most relevant magnetic instability. A dynamo model is
developed from these ingredients, and applied to the problem of angular momentum
transport in stellar interiors. It produces a predominantly horizontal field. This dynamo
process is found to be more effective in transporting angular momentum than the known 
hydrodynamic mechanisms. It might account for the observed pattern of rotation in the 
solar core.
\keywords{\emph Magnetohydrodynamics (MHD) -- stars: magnetic fields -- instabilities}
}

\maketitle

\section{Introduction}
Magnetic field generation by differential rotation is usually regarded as a process
operating in the convective zones of stars. A `convective stellar dynamo' (e.g. Parker
1979) is then a field amplification process in which differential rotation stretches field lines
into a toroidal field (i.e. an azimuthal field, with respect to the rotation axis). Convective
fluid displacements create a new poloidal field component by putting kinks into the
toroidal field lines. The stretching of these kinks produces a new toroidal component,
amplifying the existing toroidal field and thus `closing the dynamo loop'. The process is
widely believed to be responsible for the magnetic fields seen at the surface of stars with
convective envelopes like the Sun, and the study of this kind of convective field generation
has created a literature of vast proportions over the past 5 decades. 

Convection, or other imposed velocity fields like waves or shear turbulence, are not really
necessary, however, for a dynamo process to operate. An example of a magnetic field
produced by differential rotation without the assistance of an imposed small scale velocity
field is the field produced in accretion disks (Hawley et al. 1996). In this kind of
dynamo process, the differential rotation produces a small scale magnetic field on its own:
the role of an imposed velocity field in generating a new the poloidal field component is
replaced by an instability in the magnetic field {\emph itself}. The magnetic instability operates
on the toroidal field that is produced by winding up of the radial field component,
distorts it, and creates new radial field components.  The instability can either be of the
Velikhov-Chandrasekhar-Balbus-Hawley type (hereafter BH instability, see Velikhov
1959; Chandrasekhar 1960; Balbus \& Hawley 1991), or a buoyancy-driven
instability (Parker 1966), though the properties of the resulting field
may depend on which of these instabilities is the most important. 

In the same way, the generation of a magnetic field in a star requires only one essential
ingredient: a sufficiently powerful differential rotation. The recreation of poloidal field
components which is needed to close the dynamo loop can be achieved by an instability in
the toroidal field, of which there exists a variety including those (BH and buoyancy
instabilities) that operate in disks. Such instabilities do not require the presence of
convection, and can take place also in the stably stratified interior of a star. 

The instabilities of a predominantly toroidal field in stably stratified regions in a star have
been summarized from the existing literature in a previous paper (Spruit 1999). I
concluded there that the first instability likely to set in is a pinch-type instability. The
essential properties of this instability in the stellar context were established by R.J.\ Tayler
(Tayler 1973; Markey \& Tayler 1973, 1974; see also Tayler 1957; Goossens et al. 1981). What
makes it of particular importance are the absence of a threshold for instability (at least in
the absence of viscous damping and magnetic diffusion; more about this below), and the
short growth time, of the order of the Alfv\'en crossing time. It can can operate under
conditions where BH and buoyancy instabilities are suppressed by the stable
stratification. 

As an aside I note here that the solar cycle, generally considered as the classical case of a
convective dynamo process, is probably not driven by convective turbulence at all.
Properties of active regions such as the heliographic latitude and the time scale at which
they emerge, and the value of the tilt of their axes, point to field eruption from an
ordered toroidal field at the base of the convection zone. The energy density in this
toroidal field inferred from the observations is about 100 times stronger than can be
explained by convective turbulence (D'Silva \& Choudhuri 93; D'Silva and Howard
1993). The relevant processes in such a strong field are the winding-up by differential
rotation, and buoyant instability (Parker 1966) of the field itself (Caligari et al. 1998 and
references therein). This view is much closer to semiempirical models of the cycle like
those of Leighton (1969), the development of which was eclipsed by the mathematically
more interesting turbulent dynamo models.  Leighton's model is also much closer to 
the observations, since it includes the role of active region eruption as an essential ingredient 
in the operation of the dynamo process. This contrasts with the convective dynamo view, in
which active regions are only a secondary manifestation of turbulence.

\subsection{Energetics of magnetic field generation}
The differential rotation of a convectively stable stellar interior is `limited resource'.  
Whereas a dynamo operating on the differential rotation in a convective zone
feeds, indirectly, on the continuous energy flux of the star, a dynamo process in stable
layers derives its energy from the rotation of the star. This is a very small amount of
energy compared with the luminous energy, integrated over the star's life time. The
energy budget of a dynamo process in a stable zone is thus limited.

The rotational energy is available for a dynamo process only though the variation of the
rotation rate through the star. Such a gradient can result from a spin-down torque on the
star caused by angular momentum loss by a stellar wind, or from changes in the
distribution of the moment of inertia of the star as a result of  its evolution. 

\subsection{Angular momentum transport}
The main interest of a magnetic field generated in the stable layers lies in its ability to
exert internal torques. If such torques can operate on time scales short compared with the
spindown time scale or the stellar evolution time scale, respectively (depending which of
these processes is the main source of differential rotation), they can strongly affect the
degree of differential rotation that can survive in a star. In some cases, they may maintain
a state of nearly uniform rotation. For example, the usual interpretation of the rotation of
white dwarfs and neutron stars as a remnant of the initial angular momentum of their
main-sequence progenitors would require a very large degree of differential rotation
between core and envelope of these stars during their evolution on the giant and AGB
branches. Spruit \& Phinney (1998) have argued that, in the absence of more detailed
knowledge, angular momentum processes are more likely to be either too ineffective or
too effective, compared with the intermediate effectiveness needed to explain the rotation
rates of white dwarfs and neutron stars. If the process is ineffective, leading to evolution
at approximate conservation of angular momentum, the cores of stars on the AGB are
very rapidly spinning, and much too rapidly rotating white dwarfs and neutron stars are
the result. On the other hand, effective coupling leads to nearly uniform rotation and too
slow rotation of the remnants. In the latter case, other mechanisms must contribute to the
rotation of white dwarfs and neutron stars, such as birth kicks for neutron stars (Spruit
\& Phinney 1998), and nonaxisymmetric mass loss on the AGB for white dwarfs (Spruit
1998).

All known non-magnetic angular momentum transport processes in stars are weak except
at rotation near the breakup rate. If only these known processes are included,
pre-supernova cores rotate much too fast to explain the majority of supernovae (Heger et
al. 2000), though they may rotate at the right rate for the collapsar model of gamma-ray
bursts (McFadyen, Woosley \& Heger 2001 and references therein). 

Microscopic viscosity is negligible in stars as a source of internal torque. Hydrodynamic
processes such as circulation (Zahn 1992; Maeder and Zahn 1998) and hydrodynamic
instabilities are potentially  important. In practice, however, the torques transmitted by
most of these processes scale as $(\Omega/N)^2$ (where $\Omega$ is the rotation rate
and $N$ the buoyancy frequency) and hence are rather ineffective except in stars
rotating close to their breakup speed, and can not explain the low rotation rate of the 
solar core (Spruit et al. 1983). An exception is angular
momentum transport by internal waves generated by pressure fluctuations in a nearby
convective envelope (Zahn et al. 1997), which scales as a lower power of $\Omega$. Even
these waves, however, appear insufficient to explain the near-uniform rotation observed
in the solar interior (Talon \& Zahn 1998; Kumar et al. 1999).

\subsection{Evidence from the internal rotation of the Sun}
The very small degree of differential rotation in the core of the Sun, and the small
difference in rotation rate between the core and the convective envelope (Schou et al. 
1998; Charbonneau et al. 1999) requires the presence a process with two special
properties. First, the envelope is continuously spun down by the solar wind torque,
whose magnitude can be determined from in-situ measurements in the wind (Pizzo et al.
1983). The absence of an increase of rotation with depth in the core, and the
near-equality of the rotation in the core to the average rotation in the convective
envelope, require an efficient angular momentum transport process in the core. Secondly,
there is no noticeable difference of the core's rotation with latitude, in contrast with the
convective envelope, which rotates some 40\% faster at the equator than at the poles. An
isotropic angular momentum transport process coupling the core to the differentially
rotating envelope would lead to rotation with roughly equally strong vertical and
horizontal gradients. This is contrary to observation: at the transition between core and
envelope (the so-called tachocline, Spiegel \& Zahn 1992) the radial gradient is much
stronger than the horizontal gradient. A prospective angular momentum transport
process explaining this must be {\em even more} effective at reducing the gradient of
rotation on horizontal surfaces than it is in reducing the radial gradient. 

\subsection{Magnetic angular momentum transport}
Magnetic fields are quite effective at transporting angular momentum, even at strengths
much below the values observed at the surfaces of magnetically active stars or the
magnetic A stars (Mestel 1953).  The Maxwell stresses transport angular momentum in
the radial direction 
at a rate proportional to $B_rB_\phi$, and in latitude at  a rate proportional to $B_\theta
B_\phi$. If the observed internal rotation pattern in the solar core is due to magnetic
stresses, the weak gradient in latitude compared with the radial gradient then indicates
that $B_\theta\gg B_r$ in the tachocline. This fact, in combination with a sufficient amplitude
of the magnetic field, are to be explained by a prospective theory for magnetic angular
momentum transport in the core of the Sun. 

In the following I show that a differential rotation-driven dynamo operating with
magnetic  instabilities can do just this, and present a quantitative estimate of the amplitude, 
length scales and time scales of the resulting magnetic field. 

\section{The dynamo process}

I assume that the star's rotation rate is a function of the
radial coordinate $r$ only (`shellular' rotation, cf.\ Zahn 1992). This is done
for convenience only. It is quite conceivable that latitude-dependent
differential rotation, for example due to a nearby convective zone, would
have somewhat different effects.

Another assumption is that the initial magnetic field is sufficiently weak,
so that magnetic forces can be neglected initially. Thus, I ignore the
possibility that the star started with a strong magnetic field, such as those
of the magnetic A-stars or the magnetic white dwarfs.

On account of its weakness, the radial component $B_r$ of the field is wound
up by differential rotation. After only a few differential turns, the
resulting field is predominantly azimuthal, $B_\phi\gg B_r$, and its strength
increases linearly with time, until it becomes unstable. As long as the
magnetic field is weak, the energy to be gained from the field by any
instability is small, and instability is possible only for displacements that
avoid doing work against powerful restoring forces: the gas pressure and the
buoyancy force. The first instability to set in (for a review of the relevant
instabilities see Spruit 1999) is a nonaxisymmetric (typically $m=1$)
interchange in which the displacements are nearly incompressive (${\rm
div}\xi\ll \xi/r$, thus avoiding work against the gas pressure), and nearly
horizontal (along equipotential surfaces), avoiding work against the buoyancy
force. 

In the absence of diffusive damping effects such as magnetic diffusion and
viscosity, an arbitrarily weak azimuthal field is unstable (Tayler 1973; Pitts \&
Tayler 1986). The form of the instability is related to kink instabilities in a linear
pinch.  For a purely azimuthal field, the instability is a local one:
unstable displacements exist as soon as a local instability condition is
satisfied at any point in the star (Tayler 1973). Under the present
astrophysical conditions of a strongly stabilizing buoyancy force, I will refer
to pinch-type instability of an azimuthal field under stratified conditions
as they exist in a stellar interior as a Tayler instability for short.  

\subsection{time scales}
There are a number of different time scales in the problem. The fastest time
scale is the sound travel time over, say, a pressure scale height $H$,
$t_s=(g/H)^{-1/2}$, where $g$ is the acceleration of gravity. Next slower is
the buoyancy (Brunt-\-V\"ais\"al\"a-) frequency, $N=t_s^{-1}(\nabla_{\rm
ad}-\nabla)^{1/2}$, where $\nabla$ and $\nabla_{\rm ad}$ are the usual
logarithmic temperature gradient and its adiabatic value (e.g. Kippenhahn and
Weigert 1990). This frequency is of the same order as the inverse sound time
except in and close to a convective zone. Since we are dealing here
exclusively with stable, radiatively stratified zones, $N>0$ is a large
frequency. 

Next slower is the rotation rate $\Omega$ of the star. Only for rotation
rates near breakup can $\Omega$ approach $N$. The lowest frequency is the
Alfv\'en frequency,
\be \oma={B\over (4\pi\rho)^{1/2}r}. \ee
Thus, I will assume the ordering
\be \oma\ll\Omega\ll N. \ee
This limit results in significant simplifications and is the most relevant
one for the processes considered here. Under these conditions, the
characteristic growth rate of the instability is of the order
\be \sigma=\oma^2/\Om \qquad (\oma\ll\Om)\ee
(Pitts \& Tayler 1986; Spruit 1999).
For slow rotation ($\Om\ll\oma $) on the other hand, the typical growth rate
is $\oma$. The reduction by the factor $\oma/\Om$ is due to the strong
Coriolis force when rotation is rapid. This is characteristic of all
interchange-type instabilities in a rotating system.

Finally there is a time scale associated with the differential rotation:
\be t_{\Delta\Omega}=(r\partial_r\Omega)^{-1}. \ee

\subsection{Unstable displacements and instability condition}
The azimuthal field can be seen as consisting of stacks of loops concentric
with the rotation axis. The instability causes loops to slip sideways with
respect to each other by motions close to equipotential surfaces. The force
driving the instability is the magnetic pressure of the azimuthal field; the
loops in a stack press on each other with a pressure $B_\phi^2$ which can be
released by sideways displacements in a process similar to the accidental
slippage of disks in the spinal columns of bipedal vertebrates\footnote{This heuristic
picture of the instability applies close to the axis, it suggests that unstable displacements
cross the axis. The instability is actually a local one. If the instability condition is 
satisfied at any point ($r,\theta)$ in a meridional plane, there is an unstable 
displacement confined to an azimuthal annulus of some width around this point
(appendix in Tayler 1973). The region near the axis is always unstable; at
other latitudes the stability depends on the strength of the gradient of 
field strength with latitude.}.

A heuristic argument shows how the strong buoyancy force ($N\gg \oma$)
quantitatively constrains the unstable displacements. As in the case of
hydrodynamic instabilities, the effect of the buoyancy force is reduced by
thermal diffusion (Zahn 1974, 1983). On small length scales, thermal
diffusion reduces the stabilizing temperature perturbations caused by the
thermal stratification, lowering the effective buoyancy frequency. This
effect is important if the buoyancy is due mostly to the temperature
gradient, but not if the buoyancy is due mostly to the gradient in
composition. Thus we have to distinguish these two cases. Consider first the
case when the buoyancy is due to a composition gradient. The effective
buoyancy frequency is then the same on all length scales.

\subsection{Instability without thermal diffusion}

In this subsection, I ignore the thermal diffusion, but take into account the
diffusion of the magnetic field by Ohmic resistivity. For a dilute (and
non-degenerate) plasma, the magnetic diffusivity is given by Spitzer's value 
\be \eta=7\cdot 10^{11} \ln \Lambda \, T^{-3/2}. \ee

The growth rate of the instability in the absence of constraints like
rotation is $\sigma\sim\oma$. Thus, the kinetic energy released by an
unstable displacement of amplitude $\xi$ is ${1\over 2}\oma^2\xi^2$ (per unit
of mass). This energy is supplied by the field configuration. In order to
avoid wasting this energy by doing work against the stable stratification, the
unstable displacements must be nearly along equipotential surfaces,
$\xi_r\ll\xi_{\rm h}$, where $\xi_{\rm h}=\vert (\xi_\theta,\xi_\phi)\vert$
is the horizontal displacement. If $l_r$ and $l_{\rm h}$ are the radial and
horizontal length scales of the displacement, the condition that $\xi$ be
nearly incompressive implies that $l_{\rm h}/l_r\approx \xi_{\rm h}/\xi_r\gg
1$. For such displacements, the work done against the buoyancy force is (per
unit mass) ${1\over 2}\xi^2N^2(l_r/l_{\rm h})^2$. Hence the unstable
displacements have to satisfy $l_r/l_{\rm h}<\oma/N$.

In a star, the largest length scale available is of order $r$, so we must
have\footnote{An exception would be $l=1$ displacements with $\xi_r=0$,
corresponding to pure rotations of adjacent concentric spheres with respect
to each other. The radial length scale could be arbitrary for such
displacements without doing work against buoyancy. Since the field
configuration is never unstable at all latitudes, however, such displacements
probably are stable because the energy gained from the magnetic field at one
latitude is lost by work against the field at stable latitudes. This point is
not entirely certain, however, and may warrant further analysis.} $l_{\rm h}<
r$. So the radial length scale has to satisfy
\be l_r<r\oma/N. \label{lma} \ee
On the other hand, if the radial length scale is too small, field
perturbations cannot grow because the magnetic diffusivity smoothes them out
too fast. This limit is found by equating the magnetic diffusion time scale
$t_{\rm d}$ on the length scale $l_r$ to the growth time scale $\sigma^{-1}$
of the instability. In presence of a strong Coriolis force, $\Omega\gg \oma$,
the intrinsic growth rate of the instability (in the absence of
stratification and magnetic diffusion) is $\sigma\sim\oma^2/\Omega$. Hence
the radial length scale has to satisfy
\be l_r^2 > \eta\Omega/\oma^2.\label{lmi} \ee
Combining the two limits yields 
\be 
{\oma\over\Omega} >({N\over\Omega})^{1/2}({\eta\over r^2
\Omega})^{1/4}.\label{inst0}
\ee
This is, for the case when thermal diffusion can be neglected, and up to a
numerical factor of order unity, the correct instability condition as derived
from Acheson's (1978) dispersion relation for azimuthal magnetic fields
(Spruit 1999). 

\subsection{Effect of thermal diffusion}
\label{td}
Consider next the case when composition gradients can be neglected, so the
stabilizing stratification is due entirely to the entropy gradient. Since
heat is transported by photons, while viscosity and magnetic fields diffuse
by Coulomb interactions, the Prandtl number Pr$=\nu/\kappa$ and the
diffusivity ratio $s=\eta/\kappa$ are very small in a stellar interior (of
the order $10^{-6}$--$10^{-4}$). Here $\nu$ is the viscosity, $\kappa=
16\sigma T^3/(3\kappa_{\rm R}\rho^2c_p)$ is the thermal diffusivity, and
$\eta$ the magnetic diffusivity. The viscosity is of the same order but
somewhat smaller than $\eta$ so that the main mechanism  damping the
instability is magnetic diffusion. I ignore viscous damping in the following. 

If unstable displacements take place on a radial length scale $l$ and time
scale $\tau$, their temperature fluctuations diffuse away on the time scale
$\tau_{\rm T}=l^2/\kappa$. On the time scale $\tau$ they are therefore 
reduced by a factor $f=(\tau/\tau_{\rm T}+1)$. This can be taken into 
account by introducing an effective thermal buoyancy frequency $N_{\rm e}$:
\be N_{\rm e}=N/f^{1/2} \ee
The stabilizing effect of the stratification is thus reduced by thermal
diffusion, and instabilities take place more readily. This has apparently
been realized first by Townsend (1958), who used the argument in the context
of shear flows in a stratified atmosphere. It has been extended to the
astrophysical case by Zahn (1974, 1983). In the stellar evolution
community, Zahn's expression for shear instability in a low-Prandlt number
environment is sometimes referred to by the curious name `secular shear
instability'.

Applying the argument to the present case of Tayler instability, where the
stratification also plays a dominant role, we repeat the heuristic argument
in the preceding section by substituting $N_{\rm e}$ for $N$ in (\ref{lma}).
{\mk We simplify the algebra by assuming $\tau_{\rm T}\ll\tau$. This leads to an
erroneously high value for $N_{\rm e}$ in cases where $\tau_{\rm T}$ happens to be larger
than the instability time scale $\tau$. We correct for this at the end of the analysis in Sect.
\ref{discr}.} The minimum radial wavenumber $k_{\rm c}$ at which the instability can take
place is then 
\be 
(kr)_{\rm c}=({N\over\Omega})^{1/2}({r^2\Omega\over\kappa})^{1/4}. \label{kc}
\ee
The corresponding length scale $l_{\rm c}=k_{\rm c}^{-1}$ is the largest
radial length scale on which the instability can take place. Note that it is
now independent of the strength of the azimuthal field, contrary to the case
when thermal diffusion is neglected (eq \ref{lma}).
The instability condition becomes, by the same derivation as before:
\be 
{\oma\over\Omega} >({N\over\Omega})^{1/2}({\kappa\over r^2
\Omega})^{1/4}({\eta\over\kappa})^{1/2}.\label{inst1}
\ee
This is the condition derived in Spruit (1999)\footnote{Note that there is a
typo in Eq.\ (49) of that paper. The last occurrence of $N$ should be an
$\Om$. The correct result is given in the appendix of that paper in Eq.\ (A29).}. 
The conditions for validity are
\be \oma\ll N\ll\Omega, \qquad \eta\ll\kappa, \ee
and the buoyancy due to composition gradients has to be negligible (a 
different expression holds if the composition gradient dominates the
buoyancy). {\mk Since $\eta/\kappa \ll 1$, condition (\ref{inst1}) is less restrictive than
(\ref{inst0}), and instability sets in at lower field strengths. 

Both conditions (\ref{inst0}) and (\ref{inst1}), are approximate and ignore multiplying
factors of order unity. More exact conditions can be derived from Acheson's dispersion
relation (see Appendix in Spruit 1999).}

\subsection{Field generation}
\subsubsection{The field-amplification loop}
Consider the dynamo process as starting with the Tayler instability of an
azimuthal field, which itself is produced by the winding-up of a seed field.
The instability generates a new field component whose length scale in the
radial direction is small because of the strong effect of the stratification
(cf eq \ref{kc}). This instability-generated, small scale
field has zero average. The differential rotation acts on this field, winding
it up into a new contribution to the azimuthal field. This again is unstable,
thus closing the dynamo loop. Once the dynamo process has built up, the
original seed field is unimportant. All three magnetic field components then
have small length scales in the $r$-direction, but much larger scales in the
horizontal directions. 

The energy for the dynamo process is fed in by differential rotation {\emph
only}, and the small scale field is generated by instability of the azimuthal
field component itself. This is fundamentally different from a dynamo driven
by an imposed small scale velocity field such as convection. 

The process of field generation can be broken down into two conceptual steps.
In the first step, we ask ourselves at what level the Tayler instability will
saturate, if the strength of the initial azimuthal field is given. This level
determines the amplitude of the small scale processes generated by the
instability. These in turn determine the rate of decay of a given
azimuthal field by these processes. 

In the second step, we ask how fast the azimuthal field is regenerated by the
differential rotation. The process here is simply the stretching of the small
scale magnetic field by the differential rotation. The argument is then
wrapped up by requiring the regeneration rate to match the decay rate. This
yields the equilibrium field strength of the dynamo process.

\subsubsection{Analogy: convection in a stellar envelope}
These steps can be illustrated with the analogy of stellar convection. The
Rayleigh number is $Ra=gH^3(\nabla-\nabla_{\rm a})/(\nu\kappa)$, where $\nu$
and $\kappa$ are the viscosity and thermal diffusivity as before, and $g$ the
acceleration of gravity. Instability sets in when $Ra$ reaches the critical
value $Ra_{\rm c}$. The typical convective velocity and length scale $v_{\rm
c}$ and $l_{\rm c}$ can be thought of as acting like an effective diffusivity
$\nu_{\rm e}\sim v_{\rm c}l_{\rm c}$.  For $Ra\gg Ra_{\rm c}$, we now
estimate the amplitude of convection by assuming that the effective
diffusivity $\nu_{\rm e}\approx\kappa_{\rm e}$ is just so large that
$Ra=Ra_{\rm c}$, when the effective diffusivities are used in the expression
for $Ra$. In other words, the diffusivity becomes so large that the effective
Rayleigh number is just the critical value for onset of instability. With a
typical length scale $l_{\rm c}=H$, one verifies that this yields a
convective energy flux of $F_{\rm c}=\rho (gH)^{3/2}(\nabla-\nabla_{\rm
a})^{3/2}$. Up to a numerical factor of order unity, this is the mixing
length expression for the convective flux. This well-known `effective
Rayleigh number' argument is thus equivalent to a mixing length estimate.

While the previous argument yields the convective flux when the entropy
gradient $\nabla-\nabla_{\rm a}$ is given, this flux is usually fixed by
other factors. Requiring the flux to equal the value that follows from the
star's luminosity then determines the value of $\nabla-\nabla_{\rm a}$, as well as
the convective velocity amplitude $v_{\rm c}$.

We now apply the analogous steps to the magnetic dynamo process. In this
analogy, the differential rotation plays the role of the energy flux (assumed
to be imposed). The azimuthal magnetic field strength produced by winding-up
drives the instability and corresponds to the entropy gradient, while the
other magnetic field components produced by Tayler instability play the role of
the convective velocity field.

\subsubsection{Instability amplitude for given azimuthal field strength}
\label{ampl}
If the field strength is sufficiently above the critical field for Tayler
instability there is a range of radial length scales on which it can operate.
The largest of these is the length scale on which the stratification
suppresses instability, $l_{\rm c}$ (eqs \ref{kc},\ref{lma}). The smallest is
the scale on which magnetic diffusion suppresses instability. Between
these scales, the growth rate of the instability is not affected much by
either of these factors and is therefore of the order $\sigma=\oma^2/\Om$.
With the induction equation, we find that the radial field component produced
from the initial azimuthal field by the unstable displacements is
\be B_r\approx B_\phi l/r. \label{br} \ee
It grows on the instability time scale $\sigma^{-1}$, and is largest for the
maximal radial length scale $l_{\rm c}$. As in the convective example above
we now associate an effective `turbulent' diffusivity $\eta_{\rm e}$ with the
small scale field. The value of  $\eta_{\rm e}$ is then given, in the same way, by assuming
marginal stability for the given azimuthal field strength, when $\eta$ is
replaced by $\eta_{\rm e}$. From the stability conditions (\ref{inst0}) and
(\ref{inst1}) this yields
\be \eta_{\rm e0}=r^2\Om({\oma\over\Om})^4({\Om\over N})^2 \label{et0}\ee
and
\be 
\eta_{\rm e1}=r^2\Om({\oma\over\Om})^2({\Om\over N})^{1/2} ({\kappa\over
r^2N})^{1/2},\label{et1}
\ee
respectively for the case when thermal diffusion is unimportant (i.e. when
the effects of stratification is dominated by the composition gradient) and
when it dominates.

One verifies that these expressions are equivalent to setting $\eta_{\rm
e}\sim l_{\rm c}^2\sigma$, with $l_{\rm c}$ taken from (\ref{kc}) and
(\ref{lma}). As in the case of convection, the derivation in terms of
marginal stability with effective diffusivities is thus equivalent to a
simple length-and-time scale argument, but perhaps physically more
illuminating.

Because of the dominant effects of rotation and the stable stratification,
all processes are highly anisotropic. The effective diffusivity $\eta_{\rm
e}$ is to be interpreted as the relevant diffusivity for gradients in the
{\em radial} direction. 

\subsubsection{Amplitude of the dynamo-generated field}

The field strength in the above is still arbitrary. For high field strengths,
the effective diffusivity is large, since it is driven by magnetic instability, 
and thus the decay time of the magnetic field by the instability-driven
small scale processes is short. The rate at which the field is amplified, on the other hand, is
given by the imposed differential rotation. For a steady equilibrium, the damping time scale
has to match the amplification time scale. {\mk This condition sets the strength of the
magnetic field at which the process saturates.} Since we are assuming that the rotation rate
is a function of $r$ only (`shellular' rotation), only the radial field component contributes to
amplification by field line stretching. The time scale $\tau_{\rm a}$ on which $B_r$ is
amplified into an azimuthal field of the same order as the already existing $B_\phi$ is given
by
\be \tau_{\rm a} r\partial_r\Om=B_\phi/B_r. \ee
The shortest amplification time scale is obtained for the largest $B_r$, which occurs for 
the largest radial length scale on which instability takes place. Hence, with 
(\ref{br}), $B_\phi/B_r=r/l_{\rm c}$. The damping time scale at the same 
radial length scale is 
\be \tau_{\rm d}\approx l_{\rm c}^2/\eta_{\rm e}\approx
\sigma^{-1}=\Om/\oma^2.\ee
Equating the two time scales $\tau_{\rm a},\tau_{\rm d}$ yields
\be \omega_{\rm A0}/\Om=q{\Om\over N}, \label{oma0}\ee
when thermal diffusion can be ignored, and
\be 
\omega_{\rm A1}/\Om=q^{1/2}({\Om\over N})^{1/8}({\kappa\over r^2 N})^{1/8}
\label{oma1}
\ee
when it dominates. Here $q$ is the dimensionless differential rotation rate
\be q={r\partial_r\Om\over\Om}.\ee
The corresponding azimuthal and radial field strengths are
$B_{\phi}=r(4\pi\rho)^{1/2}\oma$, $B_r=B_\phi/(kr)$, hence
\be 
B_{\phi 0}=r(4\pi\rho)^{1/2}q\Om^2/N,\qquad {B_{r0}\over B_{\phi
0}}=q({\Om\over N})^2, \label{bphi0}
\ee
\be 
B_{\phi 1}=r(4\pi\rho)^{1/2}\Om q^{1/2}({\Om\over N})^{1/8}({\kappa\over r^2
N})^{1/8},
\ee
\be {B_{r1}\over B_{\phi 1}}=({\Om\over N})^{1/4}({\kappa\over
r^2N})^{1/4},\label{bphi1} 
\ee
for the two cases, respectively. {\mk These are the field strength we expect to be generated
by the dynamo process.}

\subsection{Conditions for the dynamo to operate}
In the derivation I have assumed that the differential rotation is strong
enough to maintain a dynamo process. For this to be the case, the azimuthal
field produced must be large enough for the basic ingredient, Tayler
instability, to operate. The field produced (eqs \ref{bphi0},\ref{bphi1}) can
be compared to the minimum field $B_{\rm c}$ for Tayler instability (see
\ref{inst0},\ref{inst1}):
\be 
{B_0\over B_{\rm c0}}={\oma\over\omega_{\rm Ac}}= q({\Om\over
N})^{7/4}({r^2N\over\eta})^{1/4},
\ee
\be 
{B_1\over B_{\rm c1}}=q^{1/2}({\Om\over N})^{7/8}({\kappa\over\eta})^{1/2}
({r^2N\over\kappa})^{1/8}.
\ee

Setting $B/B_{\rm c}=1$ then yields the minimal shear rate needed for the
process to operate:
\be q_0=({N\over\Om})^{7/4}({\eta\over r^2 N})^{1/4}, \label{q0}\ee
\be 
q_1=({N\over\Om})^{7/4}({\eta\over r^2 N})^{1/4} 
({\eta\over\kappa})^{3/4}.\label{q1}
\ee

\subsection{angular momentum transport}
\label{ang}
The main interest of the magnetic fields produced by the dynamo process
described lies in their ability to transport angular momentum. Since the
azimuthal field in the process is built up through the stretching of radial fields by
differential rotation, the $B_r$ and $B_\phi$ components are nearly maximally
correlated. Thus the azimuthal stress due to the field generated by the
dynamo is, for the two cases:
\be 
S_0\approx{B_{r0}B_{\phi 0}\over 4\pi}=\rho\Om^2r^2 q^3({\Om\over N})^4, \label{s0}
\ee
\be
S_1\approx\rho\Om^2r^2 q({\Om\over N})^{1/2}({\kappa\over r^2 N})^{1/2}.\label{s1}
\ee
{\mk 
Comparing these expressions, there appear to be cases where the stress evaluated with
(\ref{s1}) is {\em lower} than in case (\ref{s0}), in which the effect of thermal diffusion is
neglected. This is contrary to expectation, since thermal diffusion facilitates the instability
process. This is an artefact, resulting from the algebraic simplification made in Sect.
\ref{ampl}, where we have assumed that $\tau_{\rm T}\ll \tau$ in the expression
$f=1+\tau/\tau_{\rm T}$, so that the effective buoyancy frequency $N_{\rm e}$ is much
less than $N$. When  $\tau_{\rm T}> \tau$, the effect of thermal diffusion is absent, and
instead one should use $N_{\rm e}=N$. This can be corrected for, approximately, by
replacing (\ref{s1}) by (\ref{s0}) when the latter is larger (see below).
}

The magnetic torque on a sphere at radius $r$ is obtained from $S$ as usual by multiplying
with the lever arm $r\sin\theta$ and integrating over the sphere. The stress can also be
written in terms of an effective viscosity $\nu_{\rm e}$, through the relation 
\be S=\rho\nu_{\rm e}r\partial_r\Om \label{defnu}.\ee
 This yields:
\be \nu_{\rm e0}=r^2\Om  q^2({\Om\over N})^4,\label{nu0}\ee
\be \nu_{\rm e1}=r^2\Om ({\Om\over N})^{1/2}({\kappa\over r^2
N})^{1/2}.\label{nu1}\ee

In case 1, the effective viscosity happens to be independent of the shear rate, like a real
Newtonian viscosity. This is somewhat of a coincidence, as the comparison with case 0
shows. 

It should be stressed that the effective viscosity introduced in this way is extremely {\em
anisotropic}. Expressions (\ref{nu0},\ref{nu1}) are to be interpreted in the sense of their
definition (\ref{defnu}): they determine the {\em radial} angular momentum transport
under the assumption of `shellular' (latitude-independent) rotation. Since the magnetic
stresses are much larger in the horizontal directions than in the radial, it is likely that they
will be quite effective at wiping out horizontal gradients in $\Om$, (as suggested in fact by
the observed rotation in the solar core), so that this assumption is justified except close to a
differentially rotating convective envelope.

\section{Applications}
{\mk 
For practical application in stellar interiors, the results have to be expressed in terms of the
adiabatic buoyancy frequency $N=g[{\rm d}\rho/{\rm d}P- ({\rm d}\rho /{\rm d} P)_{\rm
ad}]^{1/2}$ and the `compositional' buoyancy frequency $N_\mu= (g{\rm d}\ln\mu/ {\rm
d}z)^{1/2}$. Define the thermal part $N_{\rm T}$ of the buoyancy frequency by $N_{\rm
T}^2=N^2-N_\mu^2$. The analysis of Sect. \ref{td} can then in principle be repeated by
using as an effective buoyancy frequency:
\be N^2_{\rm eff}=N_\mu^2+N^2_{\rm T}/(1+\tau/\tau_{\rm T}),\ee
where $\tau$ is again the instability time scale and $\tau_{\rm T}$ the thermal diffusion time
scale on length scale $l$.  This takes into account that the thermal part of the restoring force
is relaxed on long time scales and short length scales, while the compositional part can not be
reduced.}

\subsection{Effective viscosity}
{\mk
This can be carried through to obtain the generalization of expressions (\ref{s0},\ref{s1})
for the stress. The algebraic complexity of the resulting expressions, however, is
incommensurate with the level of sophistication of the present analysis. The dependences on
$N_\mu$ and $N_{\rm T}$ are probably monotonic, so that it is simpler to use a patching
formula to connect the two limiting cases. Taking also into account that the dynamo process
requires a minimum rotation gradient $q_{\min}$ to operate, a possible patching formula for
the effective radial viscosity $\nu_{r{\rm e}}$  produced by the dynamo-generated
magnetic field is:
\be \nu_{r{\rm e}}={\nu_{{\rm e}0}\nu_{{\rm e}1}\over \nu_{{\rm e}0}+\nu_{{\rm e}1}} f(q),
\label{sphi}\ee
where
\be \nu_{\rm e0}=r^2\Om  q^2({\Om\over N_\mu})^4, \label{numu}\ee
\be 
\nu_{\rm e1}=r^2\Om \max[({\Om\over N_{\rm T}})^{1/2}({\kappa\over r^2 N_{\rm
T}})^{1/2},q^2({\Om\over N_{\rm T}})^4]. \label{nut}
\ee
and
\bea 
f(q)=&1-q_{\min}/q \qquad(q>q_{\min}) \cr
               =&0 \qquad(q\leq q_{\min}). \label{fq}
\eea
The factor $f$ causes the stress to vanish smoothly as the gradient of the rotation rate
approaches the minimum value required for dynamo action. The $\max$ operator in
(\ref{nut}) corrects for the error made in deriving (\ref{s1}) for cases where thermal
diffusion has no effect, as discussed in \ref{ang}. The first factor in (\ref{sphi})
guarantees that the stabilizing effects of both the compositional and the thermal stratification
on the instability process are taken into account in the final result. 

For $q_{\rm min}$ one can use a patching formula to connect eqs (\ref{q0},\ref{q1}), for
example
\be q_{\rm min}=q_0+q_1, \label{qmi}\ee
where 
\be q_0=({N_\mu\over\Om})^{7/4}({\eta\over r^2 N_\mu})^{1/4}, \label{qmu}\ee
\be 
q_1=({N_{\rm T}\over\Om})^{7/4}({\eta\over r^2 N_{\rm T}})^{1/4} 
({\eta\over\kappa})^{3/4}.\label{qt}
\ee
The sum in (\ref{qmi}) takes into account that dynamo action is possible only when
the rotation gradient is strong enough to overcome both the stabilizing effects of the
$\mu-$gradient and of the thermal buoyancy.

\subsection{Mixing}
The fluid motions involved in the dynamo process also imply a certain amount of mixing. The 
mixing in the radial direction is effected by the same displacements that produce the effective
magnetic diffusivity. Up to a factor of unity we can set the effective diffusivity $D{\rm e}$ for
mixing in gradients of composition equal to $\eta_{\rm e}$. With (\ref{et0},\ref{et1}) and
(\ref{oma0},\ref{oma1}), patching in the same way as Eqs. (\ref{sphi}-\ref{nut}) yields for the
effective diffusivity,
\be D_{\rm e}= {D_{{\rm e}0}D_{{\rm e}1}\over D_{{\rm e}0}+D_{{\rm e}1}}f(q),\label{de} \ee
where $f$ is defined in (\ref{fq}), and
\be D_{{\rm e}0}=r^2\Om q^4({\Om\over N_\mu})^6 \label{d0}\ee
\be 
D_{{\rm e}1}=r^2\Om \max[q ({\Om\over N_{\rm T}})^{3/4}({\kappa\over r^2N_{\rm
T}})^{3/4}, q^4({\Om\over N_{\rm T}})^6]. \label{d1}
\ee
One verifies that the max operator in (\ref{d1}) switches between its arguments under the same
conditions as it does in (\ref{nut}).  

From these expressions one finds that $D_{\rm e}$ is much
smaller than $\nu_{\rm e}$, approaching it only when $r/N{\rm d}\Om/{\rm d}r$ is of order
unity. This reflects the fact that the angular momentum transport is done by magnetic stresses,
which are much more effective than the Reynolds stresses associated with the fluid motions in
the dynamo process. 
}

\section{Discussion of the result}
\label{discr}
{\mk 
The magnetic stress or its equivalent effective viscosity (\ref{sphi}- \ref{nut}) is the net
effect of several processes: the winding-up of radial fields by differential rotation, the
instability of the resulting azimuthal field, and the internal dissipation of small scale fields by
reconnection. Together these constitute a dynamo process which generates a magnetic 
field whose strength is governed by the differential rotation. Since the length scale of the 
field turns out to be small in the
radial direction, the radial transport of angular momentum by the magnetic stresses can be
represented by an effective radial viscosity [as defined in (\ref{sphi})]. Both the length
scales and the field strength itself are much larger in the horizontal directions
($\theta,\phi$). Variations in the rotation rate on horizontal surfaces are therefore likely to
be quite small.

Comparison of (\ref{numu}) and (\ref{nut}) shows that the presence of a gradient in
composition ($\mu$) has a dramatic effect. As long as this gradient can be neglected the
effective radial viscosity scales as $\Om^{3/2}$ (for fixed dimensionless differential rotation
$q$); when it dominates, the dependence is as $\Om^5$. The steep dependence on $\Om$ 
can be accounted for by noting that the stress $S$ is proportional to $B_r$ and $B_\phi$, 
while both $B_\phi$ and $B_r/B_\phi$ contain a factor $(\Om/N)^2$. The stabilizing effect
of the stratification thus effectively appears three times. 
}

{\mk
The dependence on $\Om$ is steeper than in some purely hydrodynamic transport
mechanisms. On the other hand, the scale of the effective viscosity multiplying this
dependence, of the order $r^2\Om q^2$, is much larger. To see what this implies, consider
for comparison Zahn's (1973, 1983) formalism for shear instability in the presence of thermal
diffusion. Assuming a critical Reynolds number $Re_{\rm c}\sim 10^3$ for small scale
turbulence, this process yields an effective viscosity of
\be \nu_{\rm Z}\sim\kappa, \label{nuz}\ee
while the critical shear rate for it to work is
\be q_{\rm min,Z}\sim {N\over\Om}(Re_{\rm c}{\nu\over\kappa})^{1/2}. \label{zahn}\ee
Comparing $q_{\rm min,Z}$ with (\ref{q1}), one finds that the dynamo process has the
lower value for the critical shear rate when
\be ({\Om\over N})^3> {\kappa\over  r^2 N}({\eta\over\nu Re_{\rm c}})^2.\ee
Since $\nu/\eta\sim 1$, and $\kappa/(r^2 N)$ is a very small number, the dynamo
process generates an effective viscosity already at much lower rotation rates. In the present
Sun, for example it operates, while hydrodynamic turbulence according to Zahn's formalism
is excluded by condition (\ref{zahn}). Under conditions when both hydrodynamic
turbulence according to Zahn's estimate and the dynamo process operate, the effective
viscosity of the dynamo process is larger, provided that [compare (\ref{nuz},\ref{nu1})]:
\be {\Om\over N}> ({\kappa\over r^2 N})^{1/3}.\ee
For solar parameters, $({\kappa/(r^2 N)})^{1/3}\approx 10^{-3}$, hence the effective
viscosity of the dynamo process is also larger, when both processes are possible.

The high effectiveness of the dynamo process in the absence of composition gradients is
likely to remove most of the differential rotation in compositionally homogeneous layers of a
star, with all the gradients concentrating into the inhomogeneous layers. The steep
dependence on $\Om$  will cause a certain `convergence': somewhat independent of
other factors influencing the rotation, there will be a tendency for $\Om/N_\mu$ to evolve
into a limited range. 
}

{\mk
The process as described here makes the essential assumption that the initial seed field being
wound up is sufficiently weak. For strong initial fields, such as those observed in magnetic
A-stars and white dwarfs, the process does not apply. Instead, in such stars rotation is
likely to become and remain homogenous on a short time scale (cf discussion in Spruit 1999).

Assuming that the initial field is indeed weak, it takes a while for the dynamo process to
reach its equilibrium amplitude. This may still be a sufficiently short time scale in most
circumstances. 

Since the dynamo process produces its own `turbulent diffusion', it enhances the decay of an
initial field\footnote{Strictly speaking: only the non-potential parts of this field. If the
dynamo operates only in a limited range of radii, it can not eliminate a field that is anchored
in other parts of the star}. Thus the process also enhances the distinction between two
regimes: an initially weak field serves as a seed field but is eliminated by the action of
the dynamo process, while a stronger initial field will prevent the dynamo process from
operating altogether.
}

\section{Estimates for the Sun}
For the present Sun, with a rotation rate $\Om=3\cdot 10^{-6}$, the buoyancy
frequency in the outer core, with radius $r=5\cdot  10^{10}$, is of the order
$N=10^{-3}$, while $\kappa\approx 2\cdot 10^8$, $\eta\approx 1.5\cdot 10^3$. In the
so-called tachocline (Spiegel \& Zahn 1992), the rotation rate changes from a
pattern with a variation of 30\% between equator and pole in the convection
zone, to a uniform rotation in the core. The width of this transition is
believed (Charbonneau et al. 1999) to be of the order $0.05R_\odot$, so that the 
shear rate is $q\sim 2$ in the tachocline (but varying and changing sign with 
latitude). In this region of the core, composition gradients (mostly due to the
gravitational settling of Helium) are quite small, so that case 1 applies. 
With these numbers, we find
\be {B\over B_{\rm c}}\approx 50. \ee
so that the conditions for dynamo action are satisfied. The expected field strengths
are
\be B_\phi\approx 1.5\cdot 10^4 {\rm~G},\qquad B_r\approx 1 {\rm~G}.\ee
With this field strength, the Alfv\'en frequency is $\oma\approx 10^{-7}$
s$^{-1}$. The actual frequency of Alfv\'enic modes is reduced by the strong
Coriolis force. Waves with length scale $\sim r$ have a frequency 
\be \sigma=\oma^2/\Om\sim 3\cdot 10^{-9},\ee
which is also the characteristic growth rate of the instability. This
corresponds to a time scale of the order 100 years.

\section{Discussion}
As in the case of accretion disks, an externally imposed small scale velocity field is not
necessary for obtaining field amplification by a dynamo process in stars. In addition to
differential rotation, the only ingredient required is an internal instability in a toroidal
magnetic field, to provide new poloidal field components on which differential rotation
can act. Several distinct instabilities are candidates for this key ingredient: Balbus-Hawley
instability, buoyant (Parker-) instability, and the toroidal-field instability studied by
Tayler. I have argued here and in Spruit (1999) that the latter is probably the most 
relevant, and have shown how a small-scale magnetic field may be produced in 
stably stratified layers in stars through this instability, with
differential rotation as the energy source of the process. The required differential rotation
may be produced by an external torque like a stellar wind, or by the internal evolution of
the star. The main significance of the process thus lies in its
effect on the internal rotation of the star, and for questions such as the amount of rotation
to be expected in the cores of giant stars. This in turn is critical for the interpretation of the
rotation of pulsars (Spruit \& Phinney 1998) and white dwarfs (Spruit 1998). 

The horizontal components of the magnetic field produced by this process are much
larger than the radial component, and the radial length scale of the field is much smaller
than the horizontal length scales. Because of the small radial length scale, the effects of the
microscopic (atomic) magnetic and thermal diffusion coefficients have to be included in the
analysis. Because of these effects the gradient in the rotation rate must exceed a certain
minimum for dynamo action to set in. The measured gradients in the tachocline of the
present, slowly rotating Sun are large enough, but only by a factor of 100 or so. Because
of the large ratio of horizontal to radial field strengths, the angular momentum transport
by the magnetic stresses is very anisotropic,  and far more effective at smoothing out
horizontal variations in rotation than radial variations. This is interpreted here as the main
reason for the peculiar abrupt transition from the rotation pattern in the convection zone
to that in the core of the Sun.

It must be stressed that the quantitative estimates of the dynamo process made here
ignore all multiplying factors of order unity. Experience in similar cases shows that these
factors can sometimes compound to rather large numbers. Also, I have essentially ignored
the difference between the spherical and the cylindrical radial coordinate in making
estimates. At the level of the present analysis, there is no justification for greater
quantitative detail, but the resulting uncertainty must be borne in mind. {\mk Finally, the
analysis applies only for cases where the initial magnetic field of the star is small enough. In
strong magnetic fields like those of the magnetic A-stars and white dwarfs, the dynamo
process is suppressed.}

\begin{acknowledgements}
This research was supported in part by the National Science Foundation under Grant No.\
PHY99-07949. {\mk I thank Stan Woosley and Alex Heger for their insistence on answers which 
led to the work reported here, and for extensive discussions, questions and comments. It is a
pleasure to thank Dr.\ Andr\'e Maeder for comments which led to significant expansion and
improvement of the text.}
\end{acknowledgements}

\end{document}